\documentclass[prd,floatfix,preprintnumbers,letterpaper,nofootinbib]{revtex4}
\usepackage{graphicx}
\usepackage{epsfig}
\usepackage{amsfonts}

\preprint{IGC--08/6--4}

\newcommand{\md}{{\rm d}}

\begin{document}

\title{Fermions in Loop Quantum Cosmology and the Role of Parity}
\author{Martin Bojowald}
\affiliation{Institute for Gravitation  and the Cosmos, The
  Pennsylvania State University,
104 Davey Lab, University Park, PA  16802}\author{Rupam Das} 
\affiliation{Department of Physics and Astronomy, Vanderbilt University,
Nashville, TN  37235}
\affiliation{Institute for Gravitation and the Cosmos, 
The Pennsylvania State University,
104 Davey Lab, University Park, PA  16802}

\begin{abstract}
 Fermions play a special role in homogeneous models of quantum
 cosmology because the exclusion principle prevents them from forming
 sizable matter contributions. They can thus describe the matter
 ingredients only truly microscopically and it is not possible to
 avoid strong quantum regimes by positing a large matter content.
 Moreover, possible parity violating effects are important especially
 in loop quantum cosmology whose basic object is a difference equation
 for the wave function of the universe defined on a discrete space of
 triads. The two orientations of a triad are interchanged by a parity
 transformation, which leaves the difference equation invariant for
 ordinary matter. Here, we revisit and extend loop quantum cosmology
 by introducing fermions and the gravitational torsion they imply,
 which renders the parity issue non-trivial. A treatable locally
 rotationally symmetric Bianchi model is introduced which clearly
 shows the role of parity. General wave functions cannot be
 parity-even or odd, and parity violating effects in matter influence
 the microscopic big bang transition which replaces the classical
 singularity in loop quantum cosmology.
\end{abstract}
\maketitle
\section{INTRODUCTION}
\label{sec:INTRODUCTION}

Most cosmological models --- classical or quantum --- introduce the
matter ingredients of the universe as bosonic fields, in particular
scalar ones. While this provides a good measure for the implications
of matter energy on space-time, some effects of realistic fermionic
particles may be overlooked. Especially in homogeneous models of
quantum cosmology there is an important difference between bosonic and
fermionic models: the exclusion principle forbids large matter
energies when symmetry reduction leaves only a few, finitely many
fermionic degrees of freedom. A massive universe can then be obtained
by only two possibilities: a homogeneous description with many
different fermionic species, or inhomogeneity with many local degrees
of freedom of a few species (as in \cite{FermionQC}).

Both options differ from what is modeled by large values of
homogeneous bosonic fields which rather resemble a Bose--Einstein
condensate of many identical excitations. In fact, fermion condensates
have been suggested for such a purpose, with characteristic effects
\cite{Condensate,ImmirziLambda}. This is an effective picture starting
from an inhomogeneous perspective in which fermions condense under
certain conditions, after which a symmetry reduction can be done. It
differs from a fundamental description from fermions in quantum
cosmology where constraints due to the exclusion principle cannot as
easily be avoided. Potentially fundamental mechanisms which rely on a
large amount of bosonic matter, such as bounce scenarios to avoid the
big bang singularity, have to be reanalyzed if matter is fermionic. A
truly microscopic description will then be achieved. Here, we perform
an analysis of the role of fermions in loop quantum cosmology.

Loop quantum cosmology \cite{LivRev} provides a general mechanism for
fundamental singularity resolution \cite{Sing,BSCG}. Commonly in
quantum cosmology, when volume is used as an intrinsic measure of
time, evolution must stop at the classical singularity where the
volume vanishes. In loop quantum cosmology, by contrast, the timeline
is naturally extended, first at the kinematical quantum level, by
including orientation into the basic variables: The (densitized) triad
knows about the size as well as the orientation of the universe which
make it take all real values, not just positive ones. Vanishing size
is then no longer a boundary but an interior point of minisuperspace.

What is more, even dynamically the classical singularity is removed
because the quantized Hamiltonian constraint equation uniquely extends
any wave function defined on minisuperspace across the subset of
vanishing sizes. Dynamics is dictated by a difference equation
\cite{cosmoIV,IsoCosmo} which remains regular where classical
relativity and Wheeler--DeWitt quantum cosmology would reach their
limits. Rather than being singular, the big bang transition then
appears as a place where space flips its orientation --- turning its
inside out --- while it changes from being contracting to being
expanding.\footnote{Sometimes it is suggested that this small-volume
regime is avoided altogether because wave packets may turn around in a
bounce at some minimal non-zero volume. This indeed happens for
homogeneous models containing sufficiently much kinetic energy of
matter \cite{QuantumBigBang,BouncePert}. However, this does
not appear as a general mechanism which would be valid in this form
for generic quantum states or for inhomogeneous
situations. Fundamental singularity resolution which deals with the
wave function right at vanishing volume is thus required.}
Still, the region of vanishing volume does remain special in the
underlying recurrence scheme. Some coefficients of the difference
equation can vanish at labels corresponding to zero volume, which
leads to consistency conditions implied by the dynamical law
\cite{DynIn,Essay}. This is welcome because, at least partially, it
frees one from having to pose initial values for a wave function
independently of the dynamics. The wave function of the universe is
restricted by the theory alone, relaxing the need to pick one solution
among many which could correspond to our universe.

While the set of configurations of vanishing volume is not a boundary
within the theory, in the presence of ordinary matter one may choose
to consider parity transformations as large gauge transformations which
complete the gauge group of triad rotations to all orthogonal
transformations. Then, one would restrict solutions to only those
states which are either even or odd under parity reversal. This would
essentially factor out the orientation degree of freedom introduced by
the use of triad variables, and again demote the set of vanishing
sizes to a boundary rather than an interior regime. This factoring has
indeed been assumed in recent constructions of physical Hilbert spaces
for specific isotropic models and the corresponding intuitive bounce
pictures based on \cite{APS}.

But if this is used crucially for the constructions, what happens if
more realistic matter is included which, as we know from particle
physics, cannot be parity invariant? Do properties of the specific
solutions based on the assumption of reflection symmetry depend on the
conservation of parity by matter, and if so, how reliable are the
conclusions drawn from this assumption? Only the inclusion of parity
violating terms, at least as a possibility, can provide a sufficiently
general mechanism of singularity resolution.

It may also give rise to new effects related to the role of parity
violation in the big bang transition. If this were to happen, an
intriguing new link between particle physics and quantum gravity would
result. Seeing whether this is indeed the case requires the
introduction of fermions, which is available in loop quantum gravity
\cite{FermionHiggs,QSDV,FermionHolst} (see also
\cite{LoopFermion1,LoopFermion2,LoopFermion3}). In general, however,
the parity behavior of loop quantum gravity is highly non-trivial due
to the fact that the basic variable conjugate to the densitized triad,
namely the Ashtekar--Barbero connection, is the sum of a parity-even
and a parity-odd term. It does not have a simple parity behavior and,
moreover, it appears in quantized expressions only non-linearly
through holonomies. Even in vacuum, this makes a direct demonstration
of parity invariance of loop quantum gravity --- or the lack thereof
--- very complicated \cite{FermionHolst}.

In this article, we introduce a homogeneous model which allows one to
analyze the parity behavior in a clear-cut way. At the same time, the
model is amenable to the techniques which have been proven useful for
explicit constructions of Hamiltonian constraint equations through the
difference equations of loop quantum cosmology
\cite{cosmoIV,IsoCosmo,HomCosmo}. The microscopic nature of fermions
due to the exclusion principle is explicitly realized. As we will see,
quantization of this model does not introduce unexpected parity
violations in the absence of classical parity violations. But the
inclusion of parity violating matter interactions is possible, which
can be used to illustrate the role of parity for singularity
removal. Then indeed, wave functions change under triad
reflections. The big bang transition through vanishing sizes is a
non-trivial event, which represents true local evolution in internal
time rather than merely the application of a symmetry transformation.

\section{Classical Symmetry Reduction}
\label{sec:classical symmetry reduction}

In this section, we provide the formulation of symmetry reduced
cosmological models which may have torsion due to the presence of
fermions.  We follow the symmetry reduction of torsion-free Bianchi
class A models \cite{cosmoI,HomCosmo,Spin}, combined with the
canonical formulation of gravity with fermions
\cite{FermionHiggs,QSDV,FermionAshtekar,FermionHolst}; the general
formulation without symmetry is summarized in the Appendix. Here, we
combine these research lines and explore the symmetry reduction of
gravity coupled to fermions in a first-order formalism, implying a
theory with torsion. As we will see, there are non-trivial changes in
the underlying equations, such that the analysis done here provides a
crucial consistency test of the robustness of existing models. At the
same time, we clarify the constructions of loop quantum cosmology
\cite{LivRev} from the viewpoint of some recent developments.

\subsection{Diagonalization}
\label{subsec:reduced constraints} 

Bianchi class A models constitute all homogeneous models with a
symmetry group $S$ acting freely on the space manifold $\Sigma \cong
S$ and for which standard Hamiltonian formulations exist. The symmetry
group is characterized by its structure constants $C^I_{JK}$, which
for class A models satisfy $C^{I}_{IJ}=0$ \cite{classAB} and can be
parameterized as $C^{K}_{IJ} = \epsilon^{K}_{\ \ IJ}n^{(K)}$ with
three coefficients $n^I$ which either vanish or take values $\pm
1$. Some of these models can be reduced further by imposing rotational
symmetry with one axis (where $S$ has isotropy group $F={\rm U}(1)$)
or even isotropy ($F={\rm SO}(3)$). Later in this paper we will
present a locally rotationally symmetric (LRS \cite{MacCallum}) model
with torsion in detail.

The action of a symmetry group $S$ on $\Sigma$ provides invariant
1-forms $\omega_a^I$ which are used for the reduction of
Ashtekar--Barbero variables. For each $s\in S$, they satisfy
$s^*\omega_a^I= {\rm Ad}(s)^I_J \omega_a^J$ or, in terms of the
Lie-algebra valued 1-form $\Omega_a:= \omega_a^IT_I$ with generators
$T_I$ of $S$, $s^*\Omega_a= s^{-1}\Omega_a s$.  The left invariant
1-forms then yield the decomposition
$A_a^{i}=V_0^{-1/3}\phi^{i}_{I}\omega_{a}^{I}$ of an invariant
connection with spatially constant coefficients $\phi_{I}^{i}$ (see
the Appendix of \cite{LivRev} for more details on invariant
connections). Here, we have explicitly included a factor of
$V_0=\int\md^3x |\det(\omega_a^I)|$ of the spatial coordinate volume (or
the volume of any finite region used to define the homogeneous
variables) as it will be convenient later on. A corresponding
decomposition of the densitized triad is given by
$E^{a}_{i}=V_0^{-2/3}p^{I}_{i}X^{a}_{I}$ with $X^{a}_{I}$ being
densitized left invariant vector fields dual to the 1-forms:
$\omega_a^I X^a_J=\delta^I_J|\det (\omega^K_b)|$. The symplectic
structure of the reduced model is given by
\begin{equation}
 \left\{\phi^{i}_{I},p^{J}_{j}\right\}=\gamma \kappa
\delta^{i}_{j}\delta^{J}_{I}
\end{equation}
as it follows from $(\gamma\kappa)^{-1}\int\md^3x \dot{A}_a^iE^a_i=
(\gamma\kappa)^{-1} \dot{\phi}^i_I p^I_i$.

For the purpose of loop quantization, it is useful to further reduce
the number of independent components of the invariant connection and
its conjugate momentum. In some cases, this will allow very explicit
calculations of matrix elements of the Hamiltonian constraint and the
difference equation it implies for physical states
\cite{HomCosmo}. Both the connection and the densitized triad can be
cast into diagonal form
\begin{eqnarray}
\label{diagonalconnection}
A^{i}_{a} = V_0^{-1/3}c_{(K)}\Lambda^{i}_{K}\omega_{a}^{K}\quad,\quad
E_{i}^{a} = V_0^{-2/3}
p^{(K)}\Lambda_{i}^{K}X^{a}_{K}
\end{eqnarray}
with six spatially constant coefficients $c_I$ and $p^I$ which are
considered as the only dynamical components while $\Lambda \in {\rm
SO(3)}$ is fixed up to gauge transformations.  Using the same
$\Lambda_I^i$ for $A_a^i$ and $E^a_i$ is consistent with the Gauss
constraint for diagonal torsion-free Bianchi class A models which is
then solved identically.  From the diagonal densitized triad,
moreover, we find the co-triad $e^{i}_{a} =
V_0^{-1/3}a_{(K)}\Lambda^{i}_{K}\omega_{a}^{K}$ with
$|a_1|=\sqrt{|p^2p^3/p^1|}$ and cyclic. It determines the diagonal
anisotropic spatial metric
\[
 q_{ab}= e^i_ae^i_b= V_0^{-2/3} a_{(I)}^2
 \delta_{IJ}\omega_a^I\omega_b^J= q_{IJ} \omega_a^I\omega_b^J
\]
with three independent scale factors $V_0^{-1/3}|a_I|$.

By construction, $c_I$, $p^I$ and $a_I$ are independent of coordinates
as long as the diagonalized homogeneous form is respected. In
particular in a Bianchi I model where
$\omega_a^I=\partial_ax^I=\delta_a^I$ in terms of Cartesian
coordinates $x^I$, spatial coordinates can be rescaled arbitrarily
without affecting the basic variables. However, the specific values do
depend on $V_0$ and the choice of the integration volume. Obviously,
the $V_0$-dependence is a consequence of the symmetry reduction to
homogeneity, since $V_0$ does not occur at all in an inhomogeneous
framework. Thus, the dependence has to be interpreted with care
especially after quantization where, fundamentally, the relation to
coordinates is lost. As a consequence, the role of $V_0$ cannot be
properly understood if considerations are limited to purely
homogeneous models because only the reduction from inhomogeneity shows
how $V_0$ enters; see \cite{InhomLattice} for a discussion from the
point of view of inhomogeneous states.

Note that $p^{I}$ and $a_{K}$ are allowed to take negative values to
represent different triad orientations while the orientation of
$\Lambda \in {\rm SO(3)}$ is fixed. A parity transformation then
simply implies $p^{I}\mapsto -p^{I}$ for the triad components (leaving
coordinates unchanged), while the transformation of the $c_{I}$ is in
general more complicated. In fact, we have
$A_a^i=\widetilde{\Gamma}_a^i+ \gamma K_a^i$ with the parity-even
torsion-free spin connection
\begin{equation}\label{Gamma}
\widetilde{\Gamma}_{a}^{i}=
\frac{1}{2}\epsilon^{ijk}e_{k}^{b}(2\partial_{[b}e_{a]}^{j}+
e^{c}_{j}e_{a}^{l}\partial_{b}e_{c}^{l}) 
\end{equation}
and the odd extrinsic curvature $K_a^i=K_{ab}e^b_i$. In the
torsion-free case, it follows from (\ref{Gamma}) that the homogeneous
spin connection can be expressed as $\widetilde{\Gamma}^{i}_{a} =
\widetilde{\Gamma}_{(K)}\Lambda^{i}_{K}\omega_{a}^{K}$ \cite{Spin}
with
\begin{equation}
 \widetilde{\Gamma}_{I}=\frac{1}{2}\left(\frac{a_J}{a_K}n^J+
   \frac{a_K}{a_J}n^K- \frac{a_I^2}{a_Ja_K}n^I\right)
   \quad\mbox{for indices such that }\quad \epsilon_{IJK}=1
\end{equation}
and the same $\Lambda_I^i$ as used for the densitized triad.
Similarly $K_a^i=K_{(I)}\Lambda^I_i\omega_a^i$ also with the same
$\Lambda_I^i$. Then, $c_I=\widetilde{\Gamma}_I+\gamma K_I$ does not
have a straightforward parity behavior unless $\widetilde{\Gamma}_I=0$
(as in the Bianchi I model).

The diagonalization is sufficient to capture the crucial dynamical
behavior of Bianchi models, such as the approach to a singularity. For
the quantization, it has the advantage that it reduces SU(2) to ${\rm
U}(1)^3$: holonomies of a homogeneous connection, computed along
curves generated by the invariant vector fields $X^a_I$, take the form
$h_I^{(\mu)}=\exp(\mu\phi_I^i\tau_i)$ with a real number $\mu$
depending, e.g., on the coordinate length of a curve used to compute
the holonomy. For $\phi_I^i=c_{(I)}\Lambda_I^i$, we have
\begin{equation} \label{holonomy}
 h_I^{(\mu_I)}= \exp(\mu_I c_{(I)} \Lambda_I^i\tau_i)= 
\cos\left({\textstyle\frac{1}{2}}\mu_I c_{(I)}\right)+ 2\Lambda_I^i\tau_i 
\sin\left({\textstyle\frac{1}{2}}\mu_I c_{(I)}\right)\,.
\end{equation}
While any SU(2)-holonomy along $X_I^a$ can be written in this
way,\footnote{General curves do not provide this simple form. For
instance, along $X_1^a+X_2^a$ holonomies are not of the (almost)
periodic form in $c_1$ or $c_2$ (but in $\sqrt{c_1^2+c_2^2}$). If
curves are considered which are not even straight with respect to the
given symmetry, the behavior is more complicated due to path ordering
and do not give rise to almost periodic functions
\cite{AinvinA}. However, such curves do not play a role in the
kinematical symmetry reduction, which uses the given set of $X_I^a$ to
introduce particular quantum geometries, just like classical symmetric
metrics which are used in adapted coordinates but can look complicated
in arbitrary coordinates.}  the diagonalization implies that
$\Lambda_I^i$ becomes a mere background quantity not subject to
dynamics. Thus, it is sufficient to consider only the simple commuting
exponentials $\exp(i\mu_I c_{(I)})$ to separate diagonal
connections. After a loop quantization, as we will see in detail
below, this will have the implication that a triad representation
exists, which simplifies the analysis of dynamics considerably. In
fact, triad operators will simply be $\hat{p}^I= -i\gamma\ell_{\rm
P}^2 \partial/\partial c_I$, with the Planck length $\ell_{\rm
P}=\sqrt{\kappa\hbar}$, which form a complete commuting set. Their
eigenstates
\[
 \langle c_1,c_2,c_3|\mu_1,\mu_2,\mu_3\rangle= 
\exp\left({\textstyle\frac{1}{2}}i(\mu_1c_1+\mu_2c_2+\mu_3c_3)\right)
\]
(written here in the connection representation) form an orthonormal
basis such that the coefficients in
\[
 |\psi\rangle=\sum_{\mu_1,\mu_2,\mu_3} s_{\mu_1,\mu_2,\mu_3}
 |\mu_1,\mu_2,\mu_3\rangle
\]
form the triad representation of arbitrary states. This explicit
representation, which becomes available only after diagonalization
\cite{HomCosmo}, has been the basis of all investigations so far in
homogeneous loop quantum cosmology. As we will see in this article,
arriving at such a representation is less trivial in the presence of
torsion.

\subsection{Torsion effects}

This scheme of diagonalization of the basic torsion-free gravitational
variables relies on the fact that both the connection and its
conjugate momentum can be diagonalized with the same $\Lambda_I^i$. In
other words, the su(2) valued connection and its conjugate momentum
are parallel to each other in the tangent space of the internal
symmetry group. This can be seen from the torsion-free Gauss
constraint which expressed in terms of the diagonalized variables
takes the form $p^{(I)}c_{(I)}
\epsilon_{ijk}\Lambda^{j}_{I}\Lambda^{I}_{k}=0$ and is identically
satisfied.  However, the presence of torsion via the axial fermion
current $J_{i}$, as summarized in the Appendix, enters the Gauss
constraint (\ref{smearedgc}) implying that
\begin{equation} \label{reducedGauss}
 \phi_I^j p^I_k \epsilon_{ijk}= \frac{1}{2}\sqrt{|\det
 (p^I_j)|}J_{i}\,. 
\end{equation}
For $\phi_I^i= c_{(I)}\Lambda_I^i$ and $p^I_i=p^{(I)}\Lambda^I_i$ as
above, this would only allow vanishing spatial components of the
fermion current and severely restrict the allowed models. This
situation becomes more obvious if we try to express the spin
connection including its torsion contribution as $\Gamma^{i}_{a} =
\Gamma_{(K)}\Lambda^{i}_{K}\omega_{a}^{K}$ with the same $\Lambda^I_i$
as used for the triad: One can easily verify that the partial torsion
contribution (\ref{correctedc}) to the connection cannot be expressed
as $C^{i}_{a} = C_{(K)}\Lambda^{i}_{K}\omega_{a}^{K}$ if
$J^i\not=0$. Then also the Ashtekar--Barbero connection cannot be
diagonal in the same basis. Therefore, our first result is that the
presence of torsion does not allow us to diagonalize both canonical
variables, i.e.\ the connection and the densitized triad,
simultaneously.

Moreover, fermion terms require us to use a connection ${\cal A}_a^i$
in (\ref{correctedconnection}) which carries an extra term compared to
the Ashtekar--Barbero connection, depending on the fermion current. We
then write the new diagonal variables as
\begin{equation}
 {\cal A}_a^i = V_0^{-1/3}c_{(K)}\Lambda_K^i\omega_a^K \quad,\quad E^a_i=
V_0^{-2/3} p^{(K)} T^K_i X_K^a
\end{equation}
where in general $T^I_i\not=\Lambda_I^i$. Not both $\Lambda_I^i$ and
$T^I_i$ can be fixed because partially they are determined by
dynamical fields as, e.g., per the Gauss constraint
(\ref{reducedGauss}). This has an immediate implication for the
symplectic structure because $c_I$ and $p^I$ will no longer be
canonically conjugate:
\begin{equation} \label{SympNonDiag}
\int_{\Sigma}\md^{3}x E^{a}_{i}{\cal L}_{t}A^i_a =
p^{(I)}T^{I}_{i}{\cal
L}_{t}\left(c_{(I)}\Lambda^{i}_{I}\right)=
p^{(I)}{\cal L}_{t}\left(c_{(I)}\Lambda^{i}_{I}T^{I}_{i}\right)-
c_{(I)}p^{(I)}\Lambda^{i}_{I}{\cal L}_{t}T^{I}_{i}\,.
\end{equation}
Thus, it is not $c_I$ which is conjugate to $p^I$ but
$c_{(I)}\Lambda_I^iT^{(I)}_i$. This is not a pure connection component
but depends on the relative angles between the connection direction
$\Lambda_I^i$ and the triad direction $T^I_i$ in internal space. (It
is not possible to fix both $\Lambda_I^i$ and $T^I_i$ because this
would require six parameters while the Gauss constraint allows one to
fix only three.)  Moreover, some of the angles enter the symplectic
structure as independent variables. We can, for instance, (Euler)
parameterize $T^I_i$ as the matrix
$T(\phi_I)=\exp(\phi_3T_3)\exp(\phi_2T_1)\exp(\phi_1T_3)$ using
generators $T_I$ of SO(3). Inserting this in (\ref{SympNonDiag}) shows
that the angles $\phi_I$ acquire canonical momenta given in terms of
the angles in $\Lambda_I^i$, e.g.\ $\phi_1$ being conjugate to $-{\rm
tr}((c\cdot\Lambda) (p\cdot T(\phi_1+\pi/2,\phi_2,\phi_3)))$, where
$c$ and $p$ here denote the diagonal matrices with components
$c_{I}$ and $p^{I}$, respectively. (Taking a derivative of
$T(\phi_I)$ amounts to switching sines and cosines, which is the same
as shifting an angle by $\pi/2$.)

The corresponding phase space and the constrained system defined on it
is rather involved, and so we consider a more special case which still
allows the non-trivial implications of torsion to be seen: We are
interested in the case where the presence of a fermion current is the
sole reason for anisotropy, while the 2-dimensional space transversal
to the spatial current is rotationally invariant. We can then assume
that there are bases for ${\cal A}_a^i$ and $E_i^a$, respectively,
such that
\begin{eqnarray}
\label{so3matrices}
\Lambda^{j}_{J}= {\left(\begin{array} {ccc} {1} & {0} & {0} \\
       {0} & {{\rm{cos}}{\rho}} & {-{\rm{sin}}{\rho}} \\ {0} & 
{{\rm{sin}}{\rho}} & {{\rm{cos}}{\rho}}\end{array}\right)}\quad,\quad 
T_{j}^{J}= {\left(\begin{array} {ccc} {1} & {0} & {0} \\
       {0} & {{\rm{cos}}{\phi}} & {{\rm{sin}}{\phi}} \\ {0} & 
{-{\rm{sin}}{\phi}} & {{\rm{cos}}{\phi}}\end{array}\right)}\,,
\end{eqnarray}
where $\rho$ and $\phi$ are the only non-vanishing rotation
angles. As we will demonstrate below, this allows non-trivial
solutions where the fermion current is aligned in the 1-direction.
The Liouville term in the action can then be expressed as
\begin{widetext}
\begin{eqnarray}
\label{symptransformation}
\frac{1}{\gamma\kappa}\int_{\Sigma}\md^{3}x E^{a}_{i}{\cal L}_{t}A^i_a &=&
 \frac{1}{\gamma\kappa}p^{(I)}{\cal L}_{t}\left(c_{(I)}\Lambda^{i}_{I}T^{I}_{i}\right)-
c_{(I)}p^{(I)}\Lambda^{i}_{I}{\cal L}_{t}T^{I}_{i}\nonumber\ \\
&=& \frac{1}{\gamma\kappa}\left(\dot{c}_{1}p^{1}+ {\cal L}_{t}(c_{2}{\rm{cos}}(\rho - \phi))
p^{2} + {\cal L}_{t}(c_{3}{\rm{cos}}(\rho - \phi))p^{3} - \dot{\phi} 
(c_{2}p^{2}+c_{3}p^{3}){\rm{sin}}(\rho - \phi)\right) \nonumber\ \\
&=& \frac{1}{\gamma\kappa}\left(\dot{c}_{1}p^{1}+\dot{\tilde{c}}_{2}p^{2}+ 
\dot{\tilde{c}}_{3}p^{3}  + \dot{\phi}p_{\phi}\right)\,,
\end{eqnarray}
\end{widetext}
where we introduced
\begin{equation} \label{variables}
 \tilde{c}_{2}=c_{2}{\rm{cos}}(\rho - \phi)\quad,\quad
 \tilde{c}_{3}=c_{3}{\rm{cos}}(\rho - \phi)\quad,\quad
p_{\phi}=-(c_{2}p^{2}+c_{3}p^{3}){\rm{sin}}(\rho - \phi)\,.
\end{equation}
In these components, the symplectic structure is
\begin{eqnarray}
\label{symplectic}
\left\{c_{1},p^{1}\right\}= \gamma \kappa 
\quad,\quad \left\{\tilde{c}_{2},p^{2}\right\}= \gamma \kappa 
\quad,\quad \left\{\tilde{c}_{3},p^{3}\right\}= \gamma \kappa 
\quad,\quad \left\{\phi,p_{\phi}\right\}= \gamma \kappa\, .
\end{eqnarray}
Notice that the presence of torsion at this stage introduces a new
kinematical degree of freedom $\phi$. It will be removed after solving
the Gauss constraint (\ref{reducedGauss}), which is now non-trivial.

There is a useful interpretation of the canonical variables in the
presence of torsion: We can write, e.g.,
\[
 \tilde{c}_2= c_2\cos(\rho-\phi)= c_2 \Lambda_2^iT^2_i= \phi_2^i T^2_i
\]
in terms of the general homogeneous coefficients $\phi_I^i= c_{(I)}
\Lambda_I^i$.  Since $T^I_i$ gives the direction of $E^a_i$, we can
interpret $\tilde{c}_2$ as a component 
\[
 V_0^{-1/3}\tilde{c}_2= {\cal A}_a^i E^b_i 
\frac{X_2^a\omega_b^2}{V_0^{-2/3}p^2}
\]
of the projection of ${\cal A}_a^i$ onto $E^a_i$. In the absence of
torsion, this would be a pure connection component because ${\cal
A}_a^i$ and $E^a_i$ would be parallel. With torsion, however,
$\tilde{c}_2$ is only part of an ${\cal A}_a^i$-component: Using the
expression (\ref{Asplit}), the projection removes the term
$\epsilon^i{}_{kl} e^k_aJ^l$ perpendicular to $E_k^a$ which happens to
be the torsion contribution to extrinsic curvature. Moreover, the
projection transversal to $E^a_i$ is just (half of) the variable
$p_{\phi}$ due to the sine, which thus takes a value equal to the
torsion contribution. This agrees with the solution of the Gauss
constraint (\ref{pphiJ}) below. Recall that the identification of the
torsion contribution to extrinsic curvature used in (\ref{Asplit})
cannot be completed without partially solving equations of motion. In
the projection defining $\tilde{c}_2$ and $\tilde{c}_3$, on the other
hand, no equations of motion have been used. Thus, these canonical
variables which we are naturally led to at the basic kinematical level
present torsion-free contributions without explicitly splitting off
torsion. (Something similar happens in inhomogeneous models such as
spherical symmetry \cite{SphSymmHam} or Gowdy models
\cite{EinsteinRosenQuant}. There it is spin connection contributions
that are split off by a natural definition of canonical variables
which then allows a manageable loop quantization.)

\subsection{Reduced constraints}

In terms of the diagonal variables the Gauss constraint
(\ref{smearedgc}) becomes
\begin{equation}
\label{reducedgc}
G_{i}= \frac{1}{\gamma\kappa}
\epsilon_{ijk} c_{(I)}p^{(I)} \Lambda_I^j T_k^I-\frac{1}{2} 
\sqrt{|p^1p^2p^3|} T_i^IJ_I=
-\frac{\epsilon_{i23}}{\gamma\kappa}
(c_{2}p^{2}+c_{3}p^{3})
{\rm{sin}}(\rho - \phi)-\frac{\sqrt{|p^{1}p^{2}p^{3}|}}{2}
T^{I}_{i}J_{I}= 0\,.
\end{equation}
For $i=2,3$, it thus implies $J_2=0=J_3$ while the remaining condition
\[
 \epsilon_{i23}p_{\phi} = \frac{\gamma\kappa}{2}\sqrt{|p^{1}p^{2}p^{3}|}
T^{I}_{i}J_{I}
\]
relates $J_1$ to $p_{\phi}$:
\begin{equation} \label{pphiJ}
p_{\phi} =  \frac{\gamma\kappa}{2}\sqrt{|p^{1}p^{2}p^{3}|}J_{1}=:
\frac{1}{2}\gamma\kappa{\cal J}_{1}\; ,
\end{equation}
where ${\cal J}_{i}=\xi^{\dagger}\sigma_i\xi+
\chi^{\dagger}\sigma_i\chi$ denotes the densitized axial fermion
current (which is bilinear in half-densitized fermions $\xi$ and
$\chi$).  With the choice (\ref{so3matrices}) of bases the fermion
current $J_{i}$ is aligned along the first (fixed) internal direction:
$J_2=J_3=0$. This defines a specific class of models with a
non-trivial spatial fermion current, as $J_1$ may be non-zero.

Similarly, the diffeomorphism constraint (\ref{smeardc}) can be
written as
\begin{eqnarray}
\label{reduceddc}
{\cal D}_{a}N^{a} = - c^{K}_{IJ}\phi^{i}_{K}p^{J}_{i}N^{I} = 
N^{1} (n^{2}c_{2}p^{2}+ n^{3}c_{3}p^{3}){\rm{sin}}(\rho - \phi)=0\,,
\end{eqnarray}
where $N^{a}= N^{I}X_{I}^{a}$ with $N^{I}$ constant and $C^{K}_{IJ} =
\epsilon^{K}_{\ \ IJ}n^{(K)}$ to specify different Bianchi class A
models are used. We have also imposed that the partial derivatives of
spinor fields vanish in a homogeneous model,
e.g. $\partial_{a}\psi=0$. A conclusion to be drawn from
(\ref{reducedgc}) and (\ref{reduceddc}) is that torsion is strongly
restricted in Bianchi Class A models with $n^{2}+n^{3}\neq 0$ since
this implies that $p_{\phi} = \gamma\kappa{\cal J}_{1}/2 = 0$, and
thus all spatial components of the axial vector current vanish.

Finally, the Hamiltonian constraint (\ref{hdhamiltonianconstraint}) is
\begin{widetext}
\begin{eqnarray}
\label{reducedhamiltonian}
H_{\rm Bianchi}&=& 
\frac{\kappa^{-1}}{\sqrt{|p^{1}p^{2}p^{3}|}}\left(n^{1}c_{1}p^{2}p^{3}+
n^{2}c_{2}p^{1}p^{3}{\rm{cos}}(\rho - \phi)+n^{3}c_{3}p^{2}p^{1}
{\rm{cos}}(\rho - \phi)\right)\nonumber\ \\
&&-\frac{\kappa^{-1}\gamma^{-2}}{\sqrt{|p^{1}p^{2}p^{3}|}}
\left(c_{1}p^{1}c_{2}p^{2}{\rm{cos}}(\rho - \phi)+c_{1}p^{1}c_{3}p^{3}
{\rm{cos}}(\rho - \phi)-c_{2}p^{2}c_{3}p^{3}\right)\nonumber\ \\
&&+\frac{\kappa^{-1}\gamma^{-2}(1+\gamma^2)}{\sqrt{|p^{1}p^{2}p^{3}|}}
\left((c_{1}-{\tilde{\Gamma}}_{1})p^{1}({\tilde{\Gamma}}_{2}p^{2}+
{\tilde{\Gamma}}_{3}p^{3}){\rm{cos}}(\rho - \phi)-(c_{2}-
{\tilde{\Gamma}}_{2})p^{2}{\tilde{\Gamma}}_{3}p^{3}\right)\nonumber\ \\
&&+\frac{1}{2\sqrt{|p^{1}p^{2}p^{3}|}}\left(\gamma(c_{2}p^{2}+ 
c_{3}p^{3}){\rm{sin}}(\rho - \phi){\cal J}_{1}+\theta
\left({\tilde{\Gamma}}_{1}p^{1}+({\tilde{\Gamma}}_{2}p^{2}+
{\tilde{\Gamma}}_{3}p^{3}){\rm{cos}}(\rho - \phi)\right){\cal J}^{0}
\right)\nonumber\ \\
&&+\frac{\gamma}{4\alpha}\left(n^{1}\left|\frac{p^{2}p^{3}}{p^{1}}\right|+
n^{2}\left|\frac{p^{1}p^{3}}{p^{2}}\right|+n^{3}
\left|\frac{p^{2}p^{1}}{p^{3}}\right|\right)
{\cal J}^{0}-\frac{3\gamma\kappa\theta}{16\sqrt{|p^{1}p^{2}p^{3}|}}
\left(\frac{2}{\alpha}+\frac{\gamma \theta}{1+\gamma^2}\right)
{\cal J}_{0}^2\nonumber\ \\
&&+\frac{\kappa}{16\sqrt{|p^{1}p^{2}p^{3}|}(1+\gamma^2)}
\left(2\gamma \beta \left(3-
\frac{\gamma}{\alpha}+2\gamma^{2}\right)-\theta^{2}\right){\cal J}_{1}^{2}
\end{eqnarray}
\end{widetext}
where $\alpha$, $\beta$ and $\theta$ are defined in the Appendix.  It
is important to emphasize that since $\Gamma_{a}^{i}$ is not
diagonalized in either $\Lambda_I^i$ or $T^I_i$ in the presence of
torsion, the Hamiltonian constraint in (\ref{hdhamiltonianconstraint})
expressed in terms of $\tilde{\Gamma}_{a}^{i}$ by splitting torsion
from the spin connection is essential to obtain a controlled loop
quantization as will be shown below.

\subsection{The Bianchi I LRS Model with Torsion}

If there is an isotropy group $F = {\rm U}(1)$ for the action of the
symmetry group $S$, one obtains locally rotationally symmetric (LRS)
models. Therefore, two of the diagonal components of the connection as
well as of the triad, e.g.\ the second two for definitiveness, have to
equal each other and only two degrees of freedom are left which we
choose to be $(c_{1},p^{1})$ and $(\tilde{c}_{2},p^{2})$ embedded into
the general Bianchi model by
\begin{eqnarray*}
(c_{1},\tilde{c}_{2})\mapsto (c_{1},\tilde{c}_{2},\tilde{c}_{3})=
(c_{1},\tilde{c}_{2},\tilde{c}_{2}) \ \ , \ \ (p^{1},p^{2})\mapsto 
(p^{1},p^{2},p^{3})=(p^{1},p^{2},p^{2})\,.
\end{eqnarray*}
The symplectic structure can be pulled back by this embedding
providing Poisson brackets
\begin{eqnarray}
\label{symplecticLRS}
\left\{c_{1},p^{1}\right\}= \gamma \kappa, \ \ \ \left\{\tilde{c}_{2},
p^{2}\right\}= \frac{1}{2}\gamma \kappa, \ \ \left\{\phi,{p}_{\phi}\right\}=
 \gamma \kappa
\end{eqnarray}
from (\ref{symplectic}), where $p_{\phi}$ is now ${p}_{\phi}:=
-2c_{2}p^{2}{\rm{sin}} (\rho - \phi)$. (Solutions of this symmetry
type in the presence of torsion due to spin fluids have been studied
in \cite{BITorsionInflation,BITorsion}.)

For the LRS model, the diffeomorphism and the Hamiltonian constraints,
(\ref{reduceddc}) and (\ref{reducedhamiltonian}) respectively, further
reduce to
\begin{eqnarray}
\label{reduceddc1}
{\cal D}_{a}N^{a}  = -\frac{1}{2\gamma \kappa} N^{1} (n^{2}+ n^{3})
p_{\phi}=0
\end{eqnarray}
and
\begin{widetext}
\begin{eqnarray}
\label{lrshamiltonian}
H_{\rm LRS}&=& 
\frac{\kappa^{-1}}{|p^2|\sqrt{|p^{1}|}}\left(n^{1}c_{1}\left(p^{2}
\right)^{2}+n^{2}\tilde{c}_{2}p^{1}p^{2}+n^{3}\tilde{c}_{2}p^{2}p^{1}
-\gamma^{-2}\left(2c_{1}p^{1}\tilde{c}_{2}p^{2}+\left(\tilde{c}_{2}
p^{2}\right)^{2}+\frac{1}{4}p_{\phi}^{2}\right)\right)
\nonumber\ \\
&&+\frac{\kappa^{-1}\gamma^{-2}(1+\gamma^2)}{|p^2|\sqrt{|p^{1}|}}
\left((c_{1}-{\tilde{\Gamma}}_{1})p^{1}({\tilde{\Gamma}}_{2}p^{2}+
{\tilde{\Gamma}}_{3}p^{3})\frac{2|\tilde{c}_{2}p^2|}{\sqrt{p_{\phi}^{2}+
4(\tilde{c}_{2}p^{2})^{2}}}-\left(\frac{{\rm{sgn}}(\tilde{c}_{2}p^{2})}{2}
{\sqrt{p_{\phi}^{2}+4(\tilde{c}_{2}p^{2})^{2}}}-{\tilde{\Gamma}}_{2}p^{2}
\right){\tilde{\Gamma}}_{3}p^{3}\right)\nonumber\ \\
&&-\frac{1}{2|p^2|\sqrt{|p^{1}|}}\left(\gamma p_{\phi}{\cal J}_{1}-
\theta\left({\tilde{\Gamma}}_{1}p^{1}+({\tilde{\Gamma}}_{2}p^{2}+
{\tilde{\Gamma}}_{3}p^{3})\frac{2|\tilde{c}_{2}p^2|}{\sqrt{p_{\phi}^{2}+
4(\tilde{c}_{2}p^{2})^{2}}}\right){\cal J}^{0}\right)\nonumber\ \\
&&+\frac{\gamma}{4\alpha}\left(n^{1}\frac{(p^{2})^{2}}{|p^{1}|}+
(n^{2}+n^{3})|p^{1}|\right){\cal J}^{0}-
\frac{3\gamma\kappa\theta}{16|p^2|\sqrt{|p^{1}|}}
\left(\frac{2}{\alpha}+\frac{\gamma \theta}{1+\gamma^2}\right)
{\cal J}_{0}^2\nonumber\ \\
&&+\frac{\kappa}{16|p^2|\sqrt{|p^{1}|}(1+\gamma^2)}
\left(2\gamma \beta \left(3-
\frac{\gamma}{\alpha}+2\gamma^{2}\right)-\theta^{2}\right){\cal J}_{1}^{2}\, ,
\end{eqnarray}
\end{widetext}
where we have used the definitions of $\tilde{c}_{2}$ and $p_{\phi}$ to
write
\begin{equation}\label{cos}
 \cos(\rho - \phi)=
 \frac{2|\tilde{c}_{2}p^2|}{\sqrt{p_{\phi}^{2}+
 4(\tilde{c}_{2}p^{2})^{2}}}\,.
\end{equation}

To allow a non-vanishing ${\cal J}_1$ and to be specific, we work from
now on with the Bianchi I model. Here, the diffeomorphism constraint
(\ref{reduceddc1}) vanishes identically and does not impose any
restriction on $p_{\phi}$.  This has the additional advantage that the
resulting Hamiltonian constraint will be free of terms such as
${\sqrt{p_{\phi}^{2}+4(\tilde{c}_{2}p^{2})^{2}}}$, which lack simple
quantizations. (While there are well-defined operators with this
classical limit, given that both $p_{\phi}^{2}$ and
$(\tilde{c}_{2}p^{2})^{2}$ would be mutually commuting positive
operators whose square root can be taken after summing them, not all
the operators involved have discrete spectra. Thus, it would not be
straightforward to compute explicit matrix elements of the square root
operator which would be required for the quantized Hamiltonian. Once
the square root is quantized, its inverse in (\ref{cos}) could easily
be obtained from $2\gamma\kappa p^2\cos(\rho-\phi)=
\{\sqrt{p_{\phi}^2+4(\tilde{c}_2p^2)^2},p^2\}$.)

For the Bianchi I LRS model, we then have $\widetilde{\Gamma}_I=0$ and
thus the Hamiltonian constraint is finally given by
\begin{widetext}
\begin{eqnarray}
\label{bianchi1lrshamiltonian}
H_{\rm I\,LRS}&=& 
-\frac{\kappa^{-1}\gamma^{-2}}{|p^2|\sqrt{|p^{1}|}}\left(2c_{1}p^{1}
\tilde{c}_{2}p^{2}+\left(\tilde{c}_{2}p^{2}\right)^{2}+\frac{1}{4}
p_{\phi}^{2}\right)
-\frac{\gamma}{2|p^2|\sqrt{|p^{1}|}}p_{\phi}{\cal J}_{1}
\nonumber\ \\
&&-\frac{3\gamma\kappa\theta}{16|p^2|\sqrt{|p^{1}|}}\left(\frac{2}{\alpha}+
\frac{\gamma \theta}{1+\gamma^2}\right){\cal J}_{0}^2
+\frac{\kappa}{16|p^2|\sqrt{|p^{1}|}(1+\gamma^2)}\left(2\gamma \beta \left(3-
\frac{\gamma}{\alpha}+2\gamma^{2}\right)-\theta^{2}\right){\cal J}_{1}^{2}\, .
\end{eqnarray}
\end{widetext}
This concludes the classical symmetry reduction of canonical gravity
non-minimally coupled to fermions.

\subsection{Parity behavior}

Because we are mainly concerned about the role of parity in loop
quantum cosmology, we end this section on the classical equations with
a discussion of parity invariance. As pointed out in
\cite{FermionHolst}, parity invariance in loop quantum gravity is not
guaranteed. The Ashtekar connection is a sum of a parity-even and a
parity-odd term and thus does not have a straightforward parity
behavior. This already occurs in the absence of fermions and torsion,
but is aggravated by the parity-mixing terms of torsion contributions
due to a fermion current (see (\ref{correctedc}), noting that $J_1$ is
even and $J_0$ is odd). Classically, one can explicitly split these
contributions, which essentially amounts to replacing the Ashtekar
connection with extrinsic curvature. However, a complete splitting
requires equations of motion to be used, which will not be possible
after quantization. It is then not guaranteed that quantum corrections
due to the loop quantization will preserve parity even in vacuum or in
the absence of parity-violating matter.

The model introduced here provides a clear view on parity in the
classical theory as well as after quantization, as we will see
below. One key property is that the canonical variables
(\ref{variables}) we are led to do, in hindsight, perform the
splitting into torsion-free and torsion components without using
equations of motion. Thus, in the new variables every single term in
the Hamiltonian constraint (\ref{bianchi1lrshamiltonian}) has a clear
and simple behavior under parity: Among the gravitational variables,
only $c_1$ and $p^1$ change sign under parity (reversing orientation)
while the rest remains unchanged. (Since changing the sign of $p^2$ in
an LRS model implies a reflection of both directions related by the
rotational symmetry, it is equivalent to a triad rotation and thus
mere gauge.) This is accompanied by the usual parity transformation of
the fermions present, which implies that ${\cal J}_1$ is parity
invariant while ${\cal J}_0$ changes sign as these are space and time
components of an axial vector. In particular, it is immediately clear
from (\ref{bianchi1lrshamiltonian}) that the Hamiltonian constraint is
parity invariant for free fermions. Parity violation will only result
if suitable interactions are introduced to the model, which can easily
be done by adding e.g.\ $\sqrt{-\det g}{\cal V}_{\mu}{\cal J^{\mu}}$
with the vector current ${\cal V}^{\mu}$ to the action. We will avail
ourselves of this possibility in what follows to understand the role
of parity in the loop quantized model.

\section{Quantization of the Bianchi I LRS model}

Loop quantum cosmology allows one to complete many of the
constructions of full loop quantum gravity in simplified and explicit
forms, which then provides indications toward the physical
implications of the theory. In this section, we provide a
self-contained description of anisotropic models with an emphasis on
the effects of fermions, torsion and parity.

\subsection{Quantum Kinematics}

We start with basic variables according to the Poisson structure
(\ref{symplecticLRS}). As in any loop quantization, states in the
connection representation are constructed by taking exponentials
\begin{equation}
\label{manifold}
\exp(\mu_1c_{1}\Lambda^{i}_{1}\tau_{i})\in {\rm SU}(2)\quad,
\quad\exp(\mu_2\tilde{c}_{2}\Lambda^{i}_{2}\tau_{i})\in {\rm SU}(2)\quad,
\quad \exp(ik\phi)\in {\rm U}(1)\quad \mbox{ for all }\quad 
\mu_{I}\in {\mathbb R}, 
k \in {\mathbb Z}, \Lambda_I^i \in SO(3)
\end{equation}
as they arise in holonomies. Using holonomies in the general setting
is important for a background independent basic algebra of
variables. This crucial feature is then reflected also in symmetric
models based on exponentials of connection components. The parameters
$\mu_I$ can take any real value, corresponding to evaluating
holonomies along straight edges (tangential to $X_I^a$) of arbitrary
length. The variable $\phi$, on the other hand, was introduced as a
periodic angle in (\ref{so3matrices}), such that only strictly
periodic functions ${\rm exp}(ik\phi)$ with $k\in{\mathbb Z}$ are
allowed. This unphysical degree of freedom, which we were led to
introduce due to the presence of torsion, will be removed after
solving the Gauss constraint.

Matrix elements of the exponentials in (\ref{manifold}) form a
$C^*$-algebra of (almost) periodic functions, as seen from
(\ref{holonomy}). Any function generated by this set can be written as
\begin{equation}
\label{statefunctions}
g(c_{1},\tilde{c}_{2},\phi)= \ \sum_{\mu_{1},\mu_{2},k}
\xi_{\mu_{1},\mu_{2},k}\exp\left({\textstyle\frac{1}{2}}i\mu_{1}c_{1}+
{\textstyle\frac{1}{2}}i\mu_{2}\tilde{c}_{2}+ik\phi\right)\;,
\end{equation}
with coefficients $\xi_{\mu_1,\mu_2,k}\in{\mathbb C}$, where the sum
is over finitely many $\mu_1,\mu_2 \in {\mathbb R}$ and $k \in
{\mathbb Z}$. Note that while $g(c_{1},\tilde{c}_{2},\phi)$ is almost
periodic in $c_1$ and $\tilde{c}_{2}$, it is exactly periodic in
$\phi$. This provides a complete set of continuous functions on
$\overline{{\mathbb R}}_{\rm Bohr}\times \overline{{\mathbb R}}_{\rm
Bohr} \times S^1$, where $\overline{{\mathbb R}}_{\rm Bohr}$ is the
Bohr compactification of the real line. (By definition,
$\overline{{\mathbb R}}_{\rm Bohr}$ is the compactification of
${\mathbb R}$ such that the set of all continuous functions on it is
just the set of almost periodic functions. See e.g.\ \cite{BohrWigner}
for a recent discussion of further properties.) All spaces in the
product are compact Abelian groups and carry a unique normalized Haar
measure $\md\mu(c)$ in the case of $\overline{{\mathbb R}}_{\rm
Bohr}$, where
\[
 \int f(c)\md\mu(c):= \lim_{T\to\infty}\frac{1}{2T}\int_{-T}^T f(c)\md c\,,
\]
and $\md\phi$ for $S^1$.

By Cauchy completion, we obtain the Hilbert space as a tensor product
${\cal H}_{\rm grav}= {\cal H}_{\rm Bohr}^{\otimes 2}\otimes {\cal
H}_{S^1}$ with the Hilbert spaces ${\cal H}_{\rm
Bohr}=L^2(\overline{{\mathbb R}}_{\rm Bohr},\md\mu(c))$ and ${\cal
H}_{S^1}=L^2(S^1,\md\phi)$ of square integrable functions on the Bohr
compactification of the real line and the circle,
respectively. Orthonormal bases for these spaces are given by $\langle
c |\mu\rangle=\exp(i\mu c/2)$, $\mu\in{\mathbb R}$, and $\langle
\phi|k\rangle= \exp(i k\phi)$, $k\in{\mathbb Z}$, respectively, with
\begin{eqnarray}
\label{innerproduct}
\langle \mu|\mu'\rangle=\delta_{\mu,\mu'} \quad,\quad\langle k|k'\rangle=
\delta_{k,k'}\,.
\end{eqnarray}

The configuration variables act in the obvious manner: For all $g_1$
and $g_2$ of the form (\ref{statefunctions}), we have
\begin{equation}
\label{holonomyoperator}
\left(\widehat{g}_{1}g_{2}\right)(c_{1},\tilde{c}_{2},\phi)= 
g_{1}(c_{1},\tilde{c}_{2},\phi)g_{2}(c_{1},\tilde{c}_{2},\phi)
\end{equation}
and the momentum operators are represented by
\begin{eqnarray}
\label{momentumoperators}
\hat{p}^{1}= -i\gamma \ell_{\rm P}^{2} \frac{\partial}{\partial c_{1}} 
\quad,\quad 
\hat{p}^{2}= -i\frac{\gamma \ell_{\rm P}^{2}}{2} 
\frac{\partial}{\partial \tilde{c}_{2}}
\quad \mbox{ and }\quad \hat{p}_{\phi}= -i\gamma \ell_{\rm P}^{2} 
\frac{\partial}{\partial\phi}\, ,
\end{eqnarray}
where $\ell^{2}_{\rm P} = \kappa \hbar$. (The densitized triad in
general is quantized via fluxes, i.e.\ 2-dimensional integrations over
surfaces. In a homogeneous context, however, this is not required and
densitized triad components can directly be promoted to
operators. This simple representation exists only due to our use of
variables; had we used $c_2$ instead of $\tilde{c}_2$, the operator
$\hat{p}^2$ and thus the volume operator would have been more
complicated.) Common eigenstates of all triad operators $\hat{p}^{I}$
are
\begin{equation}
\label{cylindricalfunctions}
|\mu_{1},\mu_{2},k\rangle :=\left. |\mu_{1}\right\rangle 
\otimes \left. |\mu_{2}\right\rangle \otimes \left. |k\right\rangle \,,
\end{equation}
with
\begin{eqnarray}
\label{eigenvaluesofmomentumoperators}
\hat{p}^{1}|\mu_{1},\mu_{2},k\rangle= \frac{\gamma 
\ell_{\rm P}^{2}\mu_{1}}{2}|\mu_{1},\mu_{2},k\rangle  
\quad,\quad \hat{p}^{2}|\mu_{1},\mu_{2},k\rangle= 
\frac{\gamma \ell_{\rm P}^{2}\mu_{2}}{4}|\mu_{1},\mu_{2},k\left.
\right\rangle \quad \mbox{ and }\quad \hat{p}_{\phi}
|\mu_{1},\mu_{2},k\rangle= \gamma 
\ell_{\rm P}^{2}k|\mu_{1},\mu_{2},k\rangle \, .
\end{eqnarray}
From triad operators we construct the volume operator:
\begin{equation}
\label{volumeoperator}
\hat{V}|\mu_{1},\mu_{2},k\rangle = |\hat{p}^2|
\sqrt{|\hat{p}^1|}\;|\mu_{1},\mu_{2},k\left.\right\rangle = 
\frac{\gamma^{\frac{3}{2}}\ell_{\rm P}^{3}}{4\sqrt{2}}|\mu_{2}|
\sqrt{|\mu_1|}\; |\mu_{1},\mu_{2},k\rangle\; .
\end{equation}

The full Hilbert space is a further tensor product of ${\cal H}_{\rm
grav}$ with the fermionic Hilbert space ${\cal H}_{\rm fermion}$. We
represent the latter as the space of functions $f(\Theta_{\alpha})$ of
four independent half-densitized Grassmann-valued variables
$\Theta_{\alpha}$, $\alpha=1,\ldots,4$, for the four components
contained in the fermion fields $\xi$ and $\chi$ in this order. The
fermionic momenta $\pi_{\xi}=-i\xi^{\dagger}$ and
$\pi_{\chi}=-i\chi^{\dagger}$ then give rise to components
$\overline{\Theta}_{\alpha}$ which are represented as $\hbar
\partial/\partial\Theta_{\alpha}$. In particular, for the axial
current components ${\cal J}^0=\xi^{\dagger}\xi-\chi^{\dagger}\chi$
and ${\cal J}_1= \xi^{\dagger}\sigma_1\xi+ \chi^{\dagger}\sigma_1\chi$
we have operators
\begin{eqnarray}
 \hat{\cal J}^0 &=& \hbar\frac{\partial}{\partial\Theta_1}\Theta_1+
 \hbar\frac{\partial}{\partial\Theta_2}\Theta_2-
 \hbar\frac{\partial}{\partial\Theta_3}\Theta_3-
 \hbar\frac{\partial}{\partial\Theta_4}\Theta_4\\ \hat{\cal J}_1 &=&
 \hbar\frac{\partial}{\partial\Theta_2}\Theta_1+
 \hbar\frac{\partial}{\partial\Theta_1}\Theta_2+
 \hbar\frac{\partial}{\partial\Theta_4}\Theta_3+
 \hbar\frac{\partial}{\partial\Theta_3}\Theta_4\,.
\end{eqnarray}
(The component $\hat{\cal J}_0$ is subject to ordering ambiguities
which we can ignore here.)

The currents are easy to diagonalize: Each 2-spinor copy has
eigenstates of $\frac{\partial}{\partial\Theta_2}\Theta_1+
\frac{\partial}{\partial\Theta_1}\Theta_2$ given by $f_0(\Theta)=1$
and $f^0(\Theta)=\Theta_1\Theta_2$ of eigenvalue zero,
$f_{\pm}(\Theta)=\Theta_1\pm\Theta_2$ of eigenvalue $\pm 1$. The
tensor product of both 2-spinor copies $\xi$ and $\chi$ then gives
eigenstates of eigenvalues zero, $\pm \hbar$ and $\pm 2\hbar$ for
$\hat{\cal J}_1$. The time component $\hat{{\cal J}}^0$ has the same
eigenstates.

A general state in ${\cal H}={\cal H}_{\rm grav}\otimes {\cal H}_{\rm
fermion}$ can then be written in a form using fermion dependent
coefficient functions in the triad eigenbasis
(\ref{cylindricalfunctions}):
\begin{equation}
\label{sstates}
|s\rangle = \sum_{\mu_{1},\mu_{2},k}s_{\mu_{1},\mu_{2},k}(\Theta)
|\mu_{1},\mu_{2},k\rangle\,.
\end{equation}
One can define the coefficients $s_{\mu_{1},\mu_{2},k}(\Theta)$ for
all values of $\mu_1,\mu_2\in{\mathbb R}$ and $k\in{\mathbb Z}$ in
this way. However, gauge invariance implies that the state must be
invariant under changing the sign of $\mu_2$ because this corresponds
to a triad rotation (without changing orientation). Thus, we require
$s_{\mu_{1},\mu_{2},k}(\Theta)= s_{\mu_{1},-\mu_{2},k}(\Theta)$.

The remaining sign freedom, ${\rm sgn}\mu_1$, is physical and crucial
because it determines the relative orientation of the triad. Thus, we
have a simple action
\begin{equation}
\label{parity}
s_{\mu_{1},\mu_{2},k}(\Theta_1,\Theta_2,\Theta_3,\Theta_4)
\stackrel{\widehat{\Pi}}{\longrightarrow}  
s_{-\mu_{1},\mu_{2},k}(\Theta_3,\Theta_4,\Theta_1,\Theta_2)
\end{equation}
of the parity operator $\widehat{\Pi}$ on states. For the fermion
dependence, we have represented the parity action
$\hat{\Pi}\Psi=\gamma^0\Psi$ for Dirac spinors by switching the
fermion values $\Theta_{\alpha}$ corresponding to $\xi$ and $\chi$,
respectively. This implies
\begin{equation}
 \hat{\Pi} \hat{\cal J}_0\hat{\Pi}^{-1}= -\hat{\cal J}_0 \quad,\quad
\hat{\Pi} \hat{\cal J}_1\hat{\Pi}^{-1}= \hat{\cal J}_1\,.
\end{equation}
For gravitational operators, a direct calculation shows
\begin{eqnarray}
 \hat{\Pi} \hat{p}^1\hat{\Pi}^{-1}= -\hat{p}^1 \quad&,&\quad \hat{\Pi}
\hat{p}^2\hat{\Pi}^{-1}= \hat{p}^2\\ \hat{\Pi}
\widehat{\exp(i\mu_1c_1/2)}\hat{\Pi}^{-1}=
\widehat{\exp(-i\mu_1c_1/2)} \quad&,&\quad
\hat{\Pi}\widehat{\exp(i\mu_2\tilde{c}_2/2)}\hat{\Pi}^{-1}=
\widehat{\exp(i\mu_2\tilde{c}_2/2)} 
\end{eqnarray}
as required.

Finally, we can directly solve the Gauss constraint which requires
$\hat{p}_{\phi}= \frac{1}{2}\gamma\kappa \hat{\cal J}_1$ and thus
allows us to eliminate $k$ as an independent quantity. Using the
spectra of the operators already determined, this provides solutions
with either $k=0$ or $k=\pm 1$. In the second case, there is a
non-vanishing value of the spatial axial current ${\cal J}_1$ of size
$\pm2\hbar$. The values $\pm\hbar$ for the fermion current, which do
exist as eigenvalues, are ruled out because they do not correspond to
integer $k$. Both 2-spinors present must thus have the same or
opposite $\hat{\cal J}_1$-eigenvalues, which allows them to be parity
eigenstates. The parity behavior of the full state according to
(\ref{parity}), however, is determined by the $\mu_1$-dependence,
which required the dynamics of quantum gravity coupling the triad to
fermions.

The allowed values for the current are only microscopic and may not
seem of interest to describe a macroscopic universe of large matter
content; they all vanish in the classical limit
$\hbar\to0$. Nevertheless, this provides an interesting model where
one can study the effects of fermions and parity in loop quantum
gravity. Physically, it is also clear why the matter contribution can
only be microscopic: As always in homogeneous quantum cosmological
models, each field component is reduced to a single degree of freedom
for all of space. For the fermion, this allows only one excitation per
component due to Pauli's principle. Unlike with scalar matter, one
cannot simply make the matter content large by choosing a high
``occupation'' such as a large momentum of the scalar. Significant
fermionic matter can only be included by adding more independent
spinor fields, or by introducing inhomogeneity which provides
independent field values at different points (represented by fermions
at different vertices of a spin network state in loop quantum
gravity). Rather than being a limitation, we consider this as an
important physical property of quantum cosmology in the presence of
realistic fermionic matter.

\subsection{Quantum Dynamics: The Hamiltonian Constraint}

A useful feature of the torsion-free Bianchi I model is that the
Lorentzian Hamiltonian constraint is related to the Euclidean part
simply by $H=-\gamma^{-2}H^{(E)}$ thanks to $K^i_{[a}K^j_{b]}\propto
F_{ab}^k\epsilon_{ijk}$, making use of homogeneity as well as the fact
that the spin connection vanishes. This has been used in almost all
investigations of loop quantum cosmology so far. If this relation is
not used, one can still quantize the Lorentzian constraint following
techniques of the full theory \cite{QSDI}. This results in a more
complicated constraint operator \cite{IsoCosmo}, but without crucial
differences.

However, in the presence of torsion, such a simple relationship can be
obtained only after splitting the torsion contribution from the spin
connection as shown in (\ref{hdhamiltonianconstraint}), which is now
to be quantized: even for the Bianchi I model, $\Gamma_a^i$ is no
longer zero due to torsion. Fortunately, torsion contributions to
$\Gamma_a^i$, namely $C_a^i$ in (\ref{correctedc}), are completely
determined by second class constraints. They can thus be split off and
quantized separately together with the matter terms. For the
Bianchi I LRS model, one can use a further key simplification which,
as pointed out above, allows us to project out torsion contributions
without directly computing them. All we need to do is use the new
variable $\tilde{c}_2$ instead of $c_2$. The resulting contribution to
the gravitational Hamiltonian constraint is the same as the
torsion-free one and thus can be quantized in the same way.

Mimicking the steps done in the full theory \cite{QSDI,QSDV}, one
writes curvature components ${\cal F}_{ab}^i$ as a product of (point)
holonomies $h_{I}={\rm{cos}}(\frac{1}{2}\delta_Ic_{I})+
2\Lambda^{i}_{I}\tau_{i}{\rm{sin}} (\frac{1}{2}\delta_Ic_{I})$ forming
a closed loop, whose ``edge lengths'' are denoted as $\delta_1$ and
$\delta_2$ for the two independent directions. Moreover, using 
\begin{equation}
\label{identity1}
\frac{1}{2}
\epsilon_{abc}\epsilon^{ijk}
\frac{{E}^{b}_{j}E^{c}_{k}}{\sqrt{\det (E^d_l)}}=e_{a}^{i}  
 = \frac{2}{\gamma
  \kappa}\left\{{\cal A}_{a}^{i}(x),V\right\}
\end{equation}
relevant products of triad components, including their inverse powers,
are reduced to a Poisson bracket of the general form
$h_I\{h_I^{-1},V\}$ where $V$ is the spatial volume and $h_I$ again a
holonomy. This allows one to write an operator in compact form, which
corresponds to a densely defined operator in the full theory:
\begin{equation}
 \hat{H}_{G}=-\frac{4i{\rm
sgn}(\hat{p}^1\hat{p}^2\hat{p}^3)}{\gamma^{3}\kappa \ell_{\rm
P}^{2}\delta_1\delta_2\delta_3}\sum_{IJK} \epsilon^{IJK} {\rm
tr}\left(h_{I}h_{J}h_{I}^{-1}h_{J}^{-1}h_{K}
[h_{K}^{-1},\hat{V}]\right)\,.
\end{equation}
We can now compute the product of holonomies and take the trace
explicitly, using the basic properties of Pauli matrices. We do this
directly for LRS variables with only two independent holonomies such
that $\delta_2=\delta_3$. Moreover, the sign factor is now solely
determined by ${\rm sgn}\hat{p}^1$ since $\hat{p}^2\hat{p}^3$ cannot
be negative.  This results in \cite{HomCosmo}
\begin{eqnarray}
\label{gravitationalhamiltonian}
\hat{H}_{G}
&=& -
\frac{32i{\rm sgn}(\hat{p}^1)}{\gamma^{3}\kappa \ell_{\rm P}^{2}
\delta_1\delta_2^2}
\left(2\sin({\textstyle\frac{1}{2}}\delta_1c_{1})\cos({\textstyle\frac{1}{2}}
\delta_1c_{1})
\sin({\textstyle\frac{1}{2}}\delta_2\tilde{c}_{2})
\cos({\textstyle\frac{1}{2}}\delta_2\tilde{c}_{2})\left(\sin
({\textstyle\frac{1}{2}}\delta_2\tilde{c}_{2}){\hat{V}}\cos(
{\textstyle\frac{1}{2}}\delta_2\tilde{c}_{2})-\cos
({\textstyle\frac{1}{2}}\delta_2\tilde{c}_{2}){\hat{V}}\sin(
{\textstyle\frac{1}{2}}\delta_2\tilde{c}_{2})\right)\right.
\nonumber\ \\
&&+\left.\sin^2({\textstyle\frac{1}{2}}\delta_2\tilde{c}_{2})\cos^2(
{\textstyle\frac{1}{2}}\delta_2\tilde{c}_{2})
\left(\sin({\textstyle\frac{1}{2}}\delta_1c_{1}){{\hat{V}}}\cos(
{\textstyle\frac{1}{2}}\delta_1c_{1})-
\cos({\textstyle\frac{1}{2}}\delta_1c_{1}){{\hat{V}}}\sin(
{\textstyle\frac{1}{2}}\delta_1c_{1})\right)\right)\;.
\end{eqnarray}
Because we have implicitly eliminated the torsion contributions from
holonomies by our choice of basic variables, we can directly use this
expression as it is known from torsion-free models. The torsion
contribution will then be added to the constraint operator via the
fermion current.

We emphasize that the meaning and form of the parameters $\delta_1$
and $\delta_2$ cannot be fully elucidated purely in homogeneous
models. In the absence so far of a derivation from a full,
inhomogeneous constraint (which itself is currently subject to changes
in its general form depending on ongoing developments) it appears best
to refrain from specific, heuristic arguments as to what values
they may take. (For instance, there is currently no firm basis for a
relation of those parameters to an eigenvalue of the area operator of
the full theory, as initially proposed in \cite{Bohr}.) We therefore
follow a more general route which allows whole classes of these
parameters, and confine attention to effects which are insensitive to
the specific form. To us, this seems most advisable given that it is
not just the numerical values of these parameters but even their
possible functional dependence on basic variables which remains open;
see Sec.~\ref{s:Ref} for further discussions.

In order to quantize the matter Hamiltonian, we must in particular
quantize the inverse volume $1/p^2\sqrt{|p^1|}$. Here, we use the
standard procedure \cite{QSDV}, first writing
\[
 \frac{1}{\sqrt{|\det(E^a_i)|}}=\frac{{\rm sgn}\det(e_a^i)}{6|\det(E^d_l)}
\epsilon^{abc}\epsilon_{ijk}e^{i}_{a}e^{j}_{b}e^{k}_{c} = 
\frac{36}{\gamma^3\kappa^3}{\rm sgn}\det (e_a^i) \epsilon^{abc}\epsilon_{ijk}
\{{\cal A}_a^i,V^{1/3}\}\{{\cal A}_b^j,V^{1/3}\} \{{\cal A}_c^k,V^{1/3}\}
\]
based on (\ref{identity1}), which is then quantized to
\begin{eqnarray}
\label{inversevolume}
\widehat{\left(\frac{1}{V}\right)}&=& 
\frac{144i{\rm{sgn}}
(\hat{p}^1\hat{p}^2\hat{p}^3)}{\gamma^{3}\ell_{\rm P}^{6}\delta_1
\delta_2\delta_3} \sum_{IJK}\epsilon^{IJK}
{\rm tr}\left(h_{I}[h_{I}^{-1},\hat{V}^{1/3}]h_{J}
[h_{J}^{-1},\hat{V}^{1/3}]h_{K}[h_{K}^{-1},
\hat{V}^{1/3}]\right)\nonumber\ \\
&=& -\frac{32\cdot 81 {\rm{sgn}}(\hat{p}^1)}{\gamma^{3}
\ell_{\rm P}^{6}\delta_1\delta_2^2}
\left(\sin({\textstyle\frac{1}{2}}\delta_1c_{1})
{\hat{V}}^{1/3}
\cos({\textstyle\frac{1}{2}}\delta_1c_{1})-\cos(
{\textstyle\frac{1}{2}}\delta_1c_{1})
{\hat{V}}^{1/3}{\rm{sin}}({\textstyle\frac{1}{2}}\delta_1c_{1})\right)
\nonumber\ \\ 
&&\left(\sin({\textstyle\frac{1}{2}}\delta_2\tilde{c}_{2}){\hat{V}}^{1/3}
\cos({\textstyle\frac{1}{2}}\delta_2\tilde{c}_{2})-\cos({\textstyle\frac{1}{2}}
\delta_2\tilde{c}_{2}) {\hat{V}}^{1/3}
\sin({\textstyle\frac{1}{2}}\delta_2\tilde{c}_{2})\right)^{2}\;.
\end{eqnarray}

The action of this operator as well as the Hamiltonian constraint is
easily computed using the action of ${\rm{sin}}(\frac{1}{2}\delta_1c_{1})$ and
${\rm{cos}}(\frac{1}{2}\delta_1c_{1})$ on the triad eigenstates,
\begin{eqnarray}
\label{actionofsinandcosine}
\cos({\textstyle\frac{1}{2}}\delta_1c_{1})|\mu_{1},\mu_{2},k\rangle 
&=&\frac{1}{2}( |\mu_{1}+\delta_1,\mu_{2},k\rangle + 
|\mu_{1}-\delta_1,\mu_{2},k\rangle) \nonumber\ \\
\sin({\textstyle\frac{1}{2}}\delta_1c_{1})|\mu_{1},\mu_{2},k\rangle 
&=& -\frac{1}{2}i( |\mu_{1}+\delta_1,\mu_{2},k\rangle - 
|\mu_{1}-\delta_1,\mu_{2},k\rangle)\;,
\end{eqnarray}
and the volume operator (\ref{volumeoperator}). From matrix elements
of the Hamiltonian constraint one can then write the constraint
equation $({\hat{H}}_{G}+{\hat{H}}_{\rm matter})|s\rangle = 0$ as a
difference equation for coefficients $s_{\mu_1,\mu_2,k}(\Theta)$ of
the state in the triad representation. We do this immediately on
states solving the Gauss constraint which determines $k$ in terms of
the action of $\hat{\cal J}_1$. Dropping the label $k$ on those
states, we have
\begin{eqnarray}
\label{evolutionequation}
&&2(|\mu_{2}+3\delta_2|-|\mu_{2}+\delta_2|)\left(|\mu_{1}+2\delta_1|^{1/2}
s_{\mu_{1}+2\delta_1,\mu_{2}+2\delta_2}(\Theta)-
|\mu_{1}-2\delta_1|^{1/2}s_{\mu_{1}-2\delta_1,\mu_{2}+2\delta_2}
(\Theta)\right)\nonumber\ \\ 
&&+2(|\mu_{2}-\delta_2|-|\mu_{2}-3\delta_2|)\left(|\mu_{1}-2\delta_1|^{1/2}
s_{\mu_{1}-2\delta_1,\mu_{2}-2\delta_2}(\Theta)-
|\mu_{1}+2\delta_1|^{1/2}s_{\mu_{1}+2\delta_1,\mu_{2}-2\delta_2}
(\Theta)\right)\nonumber\ \\
&&+(|\mu_{1}+\delta_1|^{1/2}-|\mu_{1}-\delta_1|^{1/2})
\left(|\mu_{2}+4\delta_2|s_{\mu_{1},\mu_{2}+4\delta_2}(\Theta)-2|\mu_{2}|
s_{\mu_{1},\mu_{2}}(\Theta)+|\mu_{2}-4\delta_2|
s_{\mu_{1},\mu_{2}-4\delta_2}(\Theta)\right)\nonumber\ \\
&=&\frac{81}{16}|\mu_1|^{1/3}|\mu_2|^{1/3}
(|\mu_{1}+\delta_1|^{1/6}-|\mu_{1}-\delta_1|^{1/6})
(|\mu_{2}+\delta_2|^{1/3}-|\mu_{2}-\delta_2|^{1/3})^{2}\nonumber\\
&&\times\left(\left(1+4\gamma^2-\frac{2\gamma \beta}{1+\gamma^2} \left(3-
\frac{\gamma}{\alpha}
+2\gamma^{2}\right)-\frac{\theta^{2}}{1+\gamma^2}\right)
\frac{{\hat{\cal J}}_{1}^{2}}{\hbar^2}
+3\gamma \theta 
\left(\frac{2}{\alpha}+\frac{\gamma \theta}{1+\gamma^2}\right)
\frac{{\hat{\cal J}}_{0}^2}{\hbar^2}\right)s_{\mu_{1},\mu_{2}}(\Theta)\;.
\end{eqnarray}

This equation is based on a non-symmetric constraint operator because
in (\ref{gravitationalhamiltonian}) we ordered all holonomy factors to
the left and kept the commutator terms with the volume operator to the
right. It is sometimes useful to have a symmetric ordering, although
this is not strictly required for constraints. (But it is required by
some methods to derive the physical Hilbert space.) There is only one
way to order the constraint symmetrically, namely by introducing
$\frac{1}{2}(\hat{H}+\hat{H}^{\dagger})$. Other possibilities have
been suggested, such as splitting the sines and cosines and writing
some to the left, others to the right of the commutator term. They
are, for instance, useful to prove self-adjointness
\cite{SelfAdFlat}. However, this corresponds to splitting the holonomy
product $h_Ih_Jh_I^{-1}h_J^{-1}$ into different factors, which cannot
be done in a general setting where there would rather be a single
holonomy $h_{\alpha}$ around a closed loop $\alpha$. The direct
symmetrization, on the other hand, is always possible and in our case
results in a difference equation
\begin{eqnarray}
\label{symmevolutionequation}
&&2\left((|\mu_{2}+3\delta_2|-|\mu_{2}+\delta_2|)|\mu_{1}+2\delta_1|^{1/2}+
(|\mu_{2}+\delta_2|-|\mu_{2}-\delta_2|)|\mu_{1}|^{1/2}\right)s_{\mu_{1}+
2\delta_{1},\mu_{2}+2\delta_{2}}(\Theta)\nonumber\ \\ 
&&-2\left((|\mu_{2}+3\delta_2|-|\mu_{2}+\delta_2|)|\mu_{1}-2\delta_1|^{1/2}+
(|\mu_{2}+\delta_2|-|\mu_{2}-\delta_2|)|\mu_{1}|^{1/2}\right)s_{\mu_{1}-
2\delta_{1},\mu_{2}+2\delta_{2}}(\Theta)\nonumber\ \\ 
&&+2\left((|\mu_{2}-\delta_2|-|\mu_{2}-3\delta_2|)|\mu_{1}-2\delta_1|^{1/2}+
(|\mu_{2}+\delta_2|-|\mu_{2}-\delta_2|)|\mu_{1}|^{1/2}\right)s_{\mu_{1}-
2\delta_{1},\mu_{2}-2\delta_{2}}(\Theta)\nonumber\\ 
&&-2\left((|\mu_{2}-\delta_2|-|\mu_{2}-3\delta_2|)|\mu_{1}+2\delta_1|^{1/2}+
(|\mu_{2}+\delta_2|-|\mu_{2}-\delta_2|)|\mu_{1}|^{1/2}\right)
s_{\mu_{1}+2\delta_{1},\mu_{2}-2\delta_{2}}(\Theta))\nonumber\ \\
&&+\left(|\mu_{1}+\delta_1|^{1/2}-|\mu_{1}-\delta_1|^{1/2}\right)
\left((|\mu_{2}|+|\mu_{2}+4\delta_1|)s_{\mu_{1},\mu_{2}+4\delta_{2}}(\Theta)
-4|\mu_{2}|s_{\mu_{1},\mu_{2}}(\Theta)+(|\mu_{2}|+|\mu_{2}-4\delta_1|)
s_{\mu_{1},\mu_{2}-4\delta_{2}}(\Theta)\right)\nonumber\ \\
&=&\frac{81}{8}|\mu_1|^{1/3}|\mu_2|^{1/3}(|\mu_{1}+\delta_1|^{1/6}-|\mu_{1}-
\delta_1|^{1/6})(|\mu_{2}+\delta_2|^{1/3}-|\mu_{2}-\delta_2|^{1/3})^{2}
\nonumber\\
&&\times\left(\left(1+4\gamma^2-\frac{2\gamma \beta}{1+\gamma^2} \left(3-
\frac{\gamma}{\alpha}
+2\gamma^{2}\right)-\frac{\theta^{2}}{1+\gamma^2}\right)
\frac{{\hat{\cal J}}_{1}^{2}}{\hbar^2}
+3\gamma \theta 
\left(\frac{2}{\alpha}+\frac{\gamma \theta}{1+\gamma^2}\right)
\frac{{\hat{\cal J}}_{0}^2}{\hbar^2}\right)s_{\mu_{1},\mu_{2}}(\Theta)\;.
\end{eqnarray}

\subsection{Lattice refinement}
\label{s:Ref}

So far, we have left the increments $\delta_1$ and $\delta_2$
unspecified. It is clear that as constants they would not influence
the recurrence behavior of the difference equation, although specific
solutions certainly depend on their values. However, in general
$\delta_1$ and $\delta_2$ may not be constant but be functions of
$\mu_1$ and $\mu_2$; this captures the way in which the discrete
structure of a state underlying spatial expansion and contraction in
loop quantum gravity is being refined dynamically
\cite{InhomLattice,CosConst,SchwarzN}: at larger $\mu_I$, an increment
of the total size by a Planck-scale amount has a weaker relative
influence on the geometry. As a consequence, $\delta_I$ decrease with
increasing spatial extensions. This can also be seen from more direct
considerations of holonomies in inhomogeneous states and how they
appear in Hamiltonian constraint operators. Since this involves the
dynamical relation between models and a full non-symmetric theory, the
precise behavior of lattice refinement has not been completely
determined. However, consequences of different behaviors can be
explored in several models. Sometimes, this is already quite
restrictive even though it is impossible to derive a unique form of
lattice refinement based solely on homogeneous models.

Non-trivial functions, such as power laws, have a much stronger
influence than constants because they make the difference equation
non-equidistant. Solutions are then more difficult to analyze and
find, even numerically (but see
\cite{RefinedNumeric,RefinementNumeric}).  Only in the special cases
where $\delta_1\propto \mu_1^{x_1}$ and $\delta_2\propto \mu_2^{x_2}$
can the equation be mapped to an equidistant one by a redefinition of
independent variables. However, such cases have been ruled out
\cite{BHIntHol} because they do not provide the correct semiclassical
behavior near a horizon of Schwarzschild black holes, whose interior
is treated as a homogeneous Kantowski--Sachs model. (The analysis in
\cite{BHIntHol} uses corrections to classical equations due to the use
of holonomies in the loop quantization, but ignores other effects such
as quantum back-reaction \cite{EffAc,Karpacz}. This type of
phenomenological equations may not capture correctly the behavior of
strong quantum regimes such as the black hole singularity. However, if
these equations do not provide the correct semiclassical behavior in
classical regimes, this cannot be corrected by the inclusion of
quantum back-reaction. The fact that some refinement schemes are ruled
out is thus a reliable feature.) In general, one has to expect
functions of the form $\delta_1(\mu_1,\mu_2)$ and
$\delta_2(\mu_1,\mu_2)$ with a non-trivial dependence on both
arguments (which may not be of power-law form).

As we will see below, a discussion of fundamental singularity
resolution only involves the recurrence near $\mu_1=0$. This is,
fortunately, insensitive to the particular refinement scheme and thus
presents a result of much wider generality than anything which applies
at larger volume where the specific refinement can be crucial.

\section{Cosmological Implications}

It follows immediately from the difference equation
(\ref{evolutionequation}) or (\ref{symmevolutionequation}) that it is
parity invariant since all its terms change sign under
(\ref{parity}). Thus, if
$s_{\mu_1,\mu_2}(\Theta_1,\Theta_2,\Theta_3,\Theta_4)$ is a
solution, so is
$s_{-\mu_1,\mu_2}(\Theta_3,\Theta_4,\Theta_1,\Theta_2)$. In
particular, any solution can be written as a combination of even and
odd solutions
$s_{\mu_1,\mu_2}(\Theta_1,\Theta_2,\Theta_3,\Theta_4)\pm
s_{-\mu_1,\mu_2}(\Theta_3,\Theta_4,\Theta_1,\Theta_2)$. This is no
longer the case if we had matter interactions violating parity, such
as a term proportional to ${\cal V}_0{\cal J}_0$. In this case, no
parity-even or odd solutions would exist. Wave functions for $\mu_1>0$
generically differ from their form for $\mu_1<0$, even though those
values are deterministically related via the difference equation. At
this stage, the precise form of parity violations in the matter system
is crucial to determine the behavior of the wave function near the
classical singularity at $\mu_1=0$.

To complete the construction, one would solve the difference equation
and determine a physical inner product on the solution space. Ideally,
one could then compute the behavior of observables of the system and
derive detailed cosmological scenarios including the role of quantum
effects. Unfortunately, such complete descriptions at an exact level
are possible only in rare, specific models. While such models are
instructive mathematically, conclusions drawn are difficult to
interpret because one could not be certain about the robustness of
results: If specific results are available only in a few special
models where exact mathematical solutions in the physical Hilbert
space can be found, there is no guarantee that they are not just the
very result only of demanding this high mathematical control.

In this context, an aspect of particular interest is the fact that
most models of loop quantum cosmology where physical Hilbert spaces
have been constructed explicitly \cite{APS,APSCurved,Hybrid}
specifically assume parity invariance in some form and make use of the
corresponding restriction of states when parity is considered as a
large gauge transformation. As we have seen here, physical states of
quantum cosmology are neither even nor odd in triad reflections if
parity violating matter is present. It may thus be misleading to treat
parity as a large gauge transformation even in cases where matter
preserves parity. Results based on this assumption may be spurious,
and one has to re-analyze the constructions of physical Hilbert spaces
without the assumption of parity invariant states. Fortunately, the
intuitive pictures of bounces which have sometimes been derived from
physical observables are insensitive to the specific construction of
the physical Hilbert space: They can be derived analytically in a
representation independent formalism based on effective equations
\cite{BouncePert,BounceCohStates}. Then, the assumption of parity as a
large gauge transformation is not necessary, and it can be dropped
without affecting the bounce result.

At a fundamental level, singularity resolution is also insensitive to
the physical Hilbert space construction and can directly be determined
using the difference equation (\ref{evolutionequation}) or
(\ref{symmevolutionequation}). (Here, it is important that {\em all}
solutions are non-singular, which then also includes physical ones.)
In general, coefficients of a difference equation of the type obtained
in loop quantum cosmology may vanish and prevent certain values of
$s_{\mu_1,\mu_2}$ from being determined in a recurrence starting from
initial values. This happens for the non-symmetric equation
(\ref{evolutionequation}) where none of the values $\psi_{0,\mu_2}$
--- right at the classical singularity --- is determined by initial
values because their coefficients in the difference equation
vanish. (The corresponding states $|0,\mu_2\rangle$ are mantic
\cite{BSCG}.)  However, for the difference equations realized such
undetermined values, if they arise, drop out completely of the
recurrence. In particular, even though values for $\mu_1=0$ remain
undetermined by initial values in the non-symmetrized version of the
equation, coefficients at $\mu_1<0$ follow deterministically from
coefficients at $\mu_1>0$.

In parity preserving models the wave function $s_{\mu_1,\mu_2}$ for
$\mu_1<0$ could simply be the mirror image of its cousin at $\mu_1>0$,
and it had to be symmetric if parity is considered a large gauge
transformation. However, if there is parity violation, the transition
through $\mu_1=0$ constitutes true evolution since values at $\mu_1<0$
must now differ from the mirror image at $\mu_1>0$. The wave function
at $\mu_1<0$ cannot be determined simply by reflection, but it has to
be derived by local evolution through all intermediate values of
$\mu_1$. In this case, the region of $\mu_1<0$ can by no means be
removed from considerations but must be considered as a physical
domain on equal footing with that at $\mu_1>0$. In particular, the
orientation-reversing big bang transition thus becomes physical and
cannot be argued away as a large gauge transformation.

For both forms of difference equations derived here, there are
consistency conditions arising due to vanishing coefficients around
$\mu_1=0$, analogous to dynamical initial conditions
\cite{DynIn,Essay}. If we evaluate any of the difference equations at
$\mu_1=0$, matter terms drop out and we obtain the universal relation
\begin{eqnarray}
 &&\left(|\mu_2+3\delta_2|- |\mu_2+\delta_2|\right)
 s_{2\delta_1,\mu_2+2\delta_2}- 
\left(|\mu_2-\delta_2|-|\mu_2-3\delta_2|\right) s_{2\delta_1,\mu_2-2\delta_2}
\nonumber\\
&=& \left(|\mu_2+3\delta_2|- |\mu_2+\delta_2|\right)
 s_{-2\delta_1,\mu_2+2\delta_2}- 
\left(|\mu_2-\delta_2|-|\mu_2-3\delta_2|\right) s_{-2\delta_1,\mu_2-2\delta_2}
\end{eqnarray}
valid for all $\mu_2$. In particular, at $\mu_2=2\delta_2$ we have
$s_{2\delta_1,4\delta_2}= s_{-2\delta_1,4\delta_2}$. At odd integer
multiples of $\mu_2=2\delta_2$, we obtain a recurrence relation which
requires $s_{2\delta_1,2(2n+1)\delta_2}=
s_{-2\delta_1,2(2n+1)\delta_2}$ for all integer $n$. 

There are thus reflection symmetry conditions which directly follow
from the dynamical law even in the presence of parity-violating
terms. (This symmetry has been observed first in the vacuum case
\cite{GenFuncKS}.)  However, evolution away from $\mu_1=\pm 1$ depends
on whether $\mu_1$ is positive or negative if parity is not
preserved. Thus, the wave function is not mirror symmetric even though
the dynamical initial condition closely ties the values $s_{\pm
2\delta_1,\mu_2}$ to each other.

\section{Conclusions}

We have introduced fermions into the framework of loop quantum
cosmology which gave rise to several non-trivial changes due to the
presence of torsion and potential parity non-invariance. We have
observed several key features which have a bearing on cosmological
scenarios and which do not arise for bosonic matter such as scalar
fields as they are used commonly in cosmological models. First, the
amount of matter is limited for each fermionic degree of freedom due
to the exclusion principle. Thus, large matter contents as they are
sometimes used to bring a quantum cosmological model into a
semiclassical regime where it may bounce more easily cannot
straightforwardly be achieved. The only possibilities are to allow
many copies of independent fermions or inhomogeneity where fermionic
components at different points will be independent. Physically, both
possibilities are quite different from having a single bosonic field
of high occupation. The methods used here may also be of interest for
a supersymmetric version of loop quantum cosmology along, e.g., the
lines of \cite{SUSYQuantCos} (see also \cite{SuperHolst}). Fermions in
quantum cosmology also play a role for decoherence
\cite{FermionDecoherence}.

This shows that it is crucial to consider the small-volume regime of a
quantum cosmological model which cannot be avoided in the absence of
much matter energy. Here, the recurrence scheme of an underlying
difference equation of loop quantum cosmology becomes essential to
determine whether the model is singular or not. As we showed, the
singularity resolution mechanism of loop quantum cosmology \cite{BSCG}
remains unchanged under the inclusion of fermionic matter even if it
violates parity. At the same time, the model we used allows us to show
that in its realm parity violations can only arise due to matter
interactions, not due to pure gravity. In other models or the full
theory, this situation may be different because the basic objects
quantized, in particular holonomies, do not transform
straightforwardly under parity. The model introduced here thus also
serves the purpose of providing one example where parity invariance of
pure gravity can be demonstrated after a loop quantization.

If one introduces parity-violating interactions, wave functions cannot
be mirror symmetric. Then, the branches at the two opposite
orientations of triads are independent of each other, and joined
through degenerate geometries by the dynamics of loop quantum
cosmology. The big bang transition now becomes a non-trivial event
where space turned its inside out in a quantum process which in
general cannot be described by an intuitive geometrical picture such
as a simple bounce.

\section*{Acknowledgements}

This work was supported in part by NSF grant PHY0653127.

\begin{appendix}
\section{The Full Constraints}
\label{subsec:full constraints}

To set the notations, the basic configuration variables in a
Lagrangian formulation of fermionic field theory are the Dirac
bi-spinor $\Psi = \left(\psi,\eta\right)^T$ and its complex conjugate
in $\overline{\Psi}=\left(\Psi^{*}\right)^{T}\gamma^{0}$ with
$\gamma^{\alpha}$ being the Minkowski signature Dirac matrices. We
note that $\psi$ and $\eta$ transform with density weight zero and are
spinors according to the fundamental representation of ${\rm
SL}\left(2,{\mathbb C} \right)$.  Then the non-minimal coupling of
gravity to fermions can be expressed by the total action composed of
the gravitational contribution $S_G$ and the matter contribution $S_F$
resulting from the fermion field:
\begin{widetext}
\begin{eqnarray}
\label{nonminimalaction}
S\left[e,\omega,\Psi\right]&=& S_{G}\left[e,\omega\right]+
S_{F}\left[e,\omega,\Psi\right]  \\ &=& 
\frac{1}{16\pi G} \int_{M}\md^{4}x \;|e|e^{\mu}_{I}e^{\nu}_{J}
P^{IJ}_{\ \ \ KL}F^{\ \ KL}_{\mu
\nu}(\omega) + \frac{1}{2}i\int_{M} \md^{4}x \;
|e|\left(\overline{\Psi}\gamma^{I}e^{\mu}_{I}\left(1-
\frac{i}{\alpha}\gamma_{5}\right)\nabla_{\mu}\Psi -
\overline{\nabla_{\mu}\Psi}\left(1-\frac{i}{\alpha}\gamma_{5}\right)
\gamma^{I}e^{\mu}_{I}\Psi\right),\nonumber
\end{eqnarray}
\end{widetext}
Here $I,J,\ldots=0,1,2,3$ denote the internal Lorentz indices and
$\mu, \nu, \ldots=0,1,2,3$ space-time indices, and $\alpha\in{\mathbb
R}$ is the parameter for non-minimal coupling.

We have expressed the gravitational action in terms of the tetrad
field $e^{\mu}_{I}$, where $e$ is its determinant and $e^{I}_{\mu}$
the inverse, using the Holst action \cite{HolstAction}. It presents a
first order formulation where the Lorentz connection, denoted by
$\omega_{\mu}^{IJ}$ and with curvature $F^{KL}_{\mu \nu}(\omega)=
2\partial_{[\mu}\omega^{IJ}_{\nu]}+
\left[\omega_{\mu},\omega_{\nu}\right]^{IJ}$, is treated as a variable
independent of the tetrad before equations of motion are imposed.
In the Holst action, we have
\begin{equation}
\label{PIJ}
P^{IJ}_{\ \ \ KL}=\delta^{[I}_{K} \delta^{J]}_{L} - \frac{1}{\gamma}
\frac{\epsilon^{IJ}_{\ \ KL}}{2}
\end{equation}
with inverse
\[
{P^{-1}_{\ \ \
IJ}}^{KL}=\frac{\gamma^{2}}{\gamma^{2}+1}\left(\delta^{[K}_{I}
\delta^{L]}_{J} + \frac{1}{\gamma} \frac{\epsilon_{IJ}^{\ \ \
KL}}{2}\right)
\]
where $\gamma$ is the Barbero--Immirzi parameter
\cite{Immirzi,AshVarReell}.  The connection appears also in the matter
part via the covariant derivative $\nabla_{\mu}$ of Dirac spinors
defined by
\begin{equation}
\label{covariantderivative}
 \nabla_{\mu}\equiv \partial_{\mu} + \frac{1}{4}\omega^{IJ}_{\mu}
 \gamma_{[I} \gamma_{J]}\quad, \ \ \ \ \
 \left[\nabla_{\mu},\nabla_{\nu}\right] = \frac{1}{4}F^{IJ}_{\mu
 \nu}\gamma_{[I} \gamma_{J]}
\end{equation}
in terms of Dirac matrices $\gamma_I$ (which will always carry an
index such that no confusion with the Barbero--Immirzi parameter should
arise).

For $\alpha\to\infty$ we have minimal coupling from the viewpoint of
the Holst action, while $\alpha=\gamma$ corresponds to minimal
coupling from the viewpoint of Einstein--Cartan theory
\cite{FermionAshtekar}. In this article, we allow all possible real
values of $\alpha$ as introduced in \cite{FermionAshtekar} (see also
\cite{FermionImmirziNonMin}). We emphasize that we have parity
invariance for all real $\alpha$, even though some torsion components
such as those written below consist of contributions of different
parity behavior.

A canonical analysis of the action yields first and second class
constraints. To summarize the result of this analysis
\cite{FermionHolst}, we use $\kappa = 8\pi G$,
$\tau_{j}=-\frac{i}{2}\sigma_{j}$ in terms of Pauli matrices,
the axial fermion current components
\begin{equation}
J^i =\psi^{\dagger}\sigma^{i}\psi+\eta^{\dagger}\sigma^{i}\eta\quad,\quad
J^{0} = \psi^{\dagger}\psi-\eta^{\dagger}\eta
\end{equation}
and the parameters
\begin{equation}
 \theta =
1-\frac{\gamma}{\alpha} \quad\mbox{ and }\quad\beta = \gamma+\frac{1}{\alpha}
\end{equation}
which are useful as a shortcut. For a consistent loop quantization,
the half-densitized $\xi:=\sqrt[4]{q}\psi$ instead of $\psi$ (and
$\chi:=\sqrt[4]{q}\eta$ instead of $\eta$) is required to be the
classical canonical variable for fermions \cite{FermionHiggs}, and
$\pi_{\xi}= -i\xi^{\dagger}$ is the conjugate momentum for $\xi$,
using the spatial metric $q_{ab}$ and its determinant $q$.

Upon solving the second class constraints, which provides the
expression
\begin{eqnarray}
\label{correctedc}
C_{a}^{i}= \frac{\gamma
  \kappa}{4(1+\gamma^{2})}\left(\theta \ \epsilon^{j}_{\
    kl}e_{a}^{k}J^{l}- \beta e_{a}^{j} J^{0}\right)
\end{eqnarray}
for the torsion contribution to the spin connection $\Gamma_a^i$, the
usual first class constraints remain: the Gauss constraint
\begin{equation}
\label{smearedgc}
G[\Lambda] := \int_{\Sigma}\md^{3}x \Lambda^i\left(
{\cal D}_{b}P^{b}_{i}
-{\textstyle\frac{1}{2}}\sqrt{q} J_{i}\right)= 
\int_{\Sigma}\md^{3}x \Lambda^i\left({\cal D}_{b}P^{b}_{i} 
-\pi_{\xi}\tau_{i}\xi- \pi_{\chi}\tau_{i}\chi\right)\,,
\end{equation}
the diffeomorphism constraint
\begin{widetext}
\begin{eqnarray}
\label{smeardc}
D[N^a] &:=& \ \int_{\Sigma} \md^{3}x \ N^{a} \left(2
P^{b}_{j}\partial_{[a}{\cal A}_{b]}^{j}- {\cal
A}_{a}^{i}\partial_{b}P^{b}_{i}+
\frac{1}{2}(\pi_{\xi}{\partial}_{a}{\xi}-(\partial_a\pi_{\xi})\xi
+\pi_{\chi}{\partial}_{a}{\chi}- (\partial_a\pi_{\chi}) \chi)\right)\,,
\end{eqnarray}
\end{widetext}
and the Hamiltonian constraint (modulo Gauss constraint)
\begin{widetext}
\begin{eqnarray}
\label{hdhamiltonianconstraint}
H_{\rm total}[N]&=& \int_{\Sigma_{t}} \md^{3}x \ N\left(
\frac{\gamma^{2}\kappa}{2\sqrt{q}}{P}^{a}_{i}P^{b}_{j}\left(\epsilon^{ij}_{\
  \ k}{\cal F}_{ab}^{k}-2(\gamma^{2}+1){\cal K}_{[a}^{i}{\cal K}_{b]}^{ j}
\right)
+\frac{\gamma^{2}\kappa
   P^{a}_{i}}{\sqrt{q}}{\cal
   D}_{a}{\cal J}^{i}\right.+ \frac{\gamma\kappa}{2\sqrt{q}}\theta P^{a}_{i}\widetilde{\Gamma}^{i}_{a}{\cal J}^0\nonumber\\
&&-i\frac{\gamma \kappa P^{a}_{i}}{\sqrt{q}} 
\left(\pi_{\xi}\tau^{i}{\partial}_{a}\xi -\pi_{\chi}\tau^{i}{\partial}_{a}\chi 
- c.c.\right)+\frac{\gamma^{3}\kappa^{2}}{4\alpha q}\epsilon^{ij}_{\ \ k}P^{a}_{i}e^{k}_{b}{\cal J}^0\partial_{a}P^{b}_{j}
-\frac{3\gamma\kappa\theta}{16\sqrt{q}}\left(\frac{2}{\alpha}+\frac{\gamma \theta}{1+\gamma^2}\right){\cal J}_0^2\nonumber\ \\
&&+\left.\frac{\kappa}{16\sqrt{q}(1+\gamma^2)}\left(2\gamma \beta \left(3-
\frac{\gamma}{\alpha}+2\gamma^{2}\right)-\theta^{2}\right){\cal J}_{l}{\cal J}^{l}\right)
\end{eqnarray}
\end{widetext}
expressed in terms of the densitized axial fermion current components
\begin{equation}
 {\cal J}^i=\xi^{\dagger}\sigma^i\xi+ \chi^{\dagger}\sigma^i\chi \quad,\quad
{\cal J}^0= \xi^{\dagger}\xi-\chi^{\dagger}\chi\,.
\end{equation}
The Hamiltonian constraint has been written in a useful form after
splitting the spin connection $\Gamma_a^i=\widetilde{\Gamma}^{i}_{a}+
C_{a}^{i}$ with the torsion-free connection
$\widetilde{\Gamma}^{i}_{a}$.  As usually, $N$ and $N^{a}$ are,
respectively, the lapse function and shift vector used to foliate the
space-time manifold $M$.

As basic gravitational variables we use a canonical pair given by the
densitized triad $P^{a}_{i}=E^a_i/\gamma\kappa$ together with the
connection
\begin{eqnarray}
\label{correctedconnection}
{\cal A}_{a}^{i}\ := \ A_{a}^{i} + \frac{\gamma \kappa}{4\alpha}e_{a}^{i}
J^{0} \ =\widetilde{\Gamma}_{a}^{i} + {\cal C}_{a}^{i}+\gamma K_{a}^{i}
\end{eqnarray}
where
\begin{eqnarray}
{\cal C}_{a}^{i}= C_a^i+  \frac{\kappa\gamma}{4\alpha} e^i_a J^0
 =\frac{\theta\gamma^{2} \kappa}{4(1+\gamma^{2})}
  \left(\frac{1}{\gamma} \epsilon^{j}_{\ kl}e_{a}^{k}J^{l}- e_{a}^{j}
  J^{0}\right)
\end{eqnarray}
whose curvature and covariant derivative we denote as ${\cal
F}_{ab}^{k}$ and ${\cal D}$, respectively. The $J^0$-term in the
connection ${\cal A}_a^i$, compared to the Ashtekar--Barbero
connection, is required for a formulation in terms of half-densitized
fermions. Solutions of the second class constraints tell us only what
the torsion contribution $C_a^i$ to the spin connection is. To know
the torsion contribution to extrinsic curvature $K_a^i$, and thus to
the Ashtekar--Barbero connection, one has to partially solve equations
of motion. Doing so \cite{FermionHolst}, the $J^0$-term is cancelled
and we can write
\begin{equation} \label{Asplit}
 {\cal A}_a^i = \widetilde{\Gamma}_a^i+ \gamma \widetilde{K}_a^i+ 
\frac{\kappa\gamma}{4} \epsilon^i{}_{kl} e^k_aJ^l\,.
\end{equation}

\end{appendix}


\end{document}